\documentclass{emulateapj}
\usepackage{aas_macros}
\usepackage{epstopdf}
\usepackage{epsfig}
\usepackage{natbib}
\usepackage{float}
\usepackage{gensymb}
\usepackage{amsmath}
\usepackage{lipsum}
\usepackage{times}
\usepackage{comment}
\usepackage{color}
\usepackage{multirow}
\usepackage{mwe}
\usepackage{CJKutf8}            

\definecolor{purple}{RGB}{160,32,240}

\begin{document}

\title{On The Effect of Environment on Line Emission from the Circumgalactic Medium}

\author{Huanian Zhang \begin{CJK*}{UTF8}{gkai}(张华年)\end{CJK*}, Dennis Zaritsky,  Peter Behroozi \altaffilmark{1}, and Jessica Werk \altaffilmark{2}}
\altaffiltext{1}{Steward Observatory, University of Arizona, Tucson, AZ 85719, USA; fantasyzhn@email.arizona.edu}
\altaffiltext{2}{Department of Astronomy, University of Washington, Seattle, WA 98195,  USA}

\begin{abstract}
We measure differences in the emission line flux from the circumgalactic medium, CGM, of galaxies in different environments. Such differences could be a critical clue in explaining a range of galaxy properties that depend on environment. Using large samples of stacked archival spectra from the Sloan Digital Sky Survey, we find that the H$\alpha$ + [N {\small II}] emission line flux from the CGM within 50 kpc of $\sim$ L$^*$ galaxies is lower both for galaxies that lie within a projected distance of $\sim$ 500 kpc from a massive ($M_*>10^{11} M_\odot$) galaxy and for galaxies in richer/denser environments. The environmental differences  are statistically significant even  after we control for galaxy mass and morphology. We interpret these observations as a direct signature of environmentally-caused strangulation. We present a simple, heuristic model for the effect of a massive parent galaxy. In this model, the CGM cool gas fraction within 50 kpc is significantly decreased for galaxies that lie within 700 kpc of a massive galaxy, with about 80\% of the cool gas removed even when the galaxy is at a distance of 500 kpc from its massive parent. However, we discuss alternative physical causes for the observed behavior and discuss ways forward in addressing open questions. 
\end{abstract}


\section{Introduction}

Galaxy properties correlate strongly with their surrounding environment \citep{Hubble1931}. Those in rich/dense environments tend to be red and elliptical, while those that are isolated or in poor/low-density environments tend to be blue and disky \citep{Dressler1980,Postman1984,Kauffmann2004}. Not only does the surrounding environment apparently influence the properties of the luminous central components of galaxies, it also appears to affect the gaseous halos of galaxies, as inferred from C IV absorption studies of galaxies' circumgalactic media  \citep{Burchett2016}.  The importance of the circumgalactic medium, or CGM, to our  understanding of galaxy evolution comes from the fact that the CGM is the fuel reservoir for subsequent star formation in galaxies \citep{spitzer}, the depository for outflowing material, and the 
site of the majority of the baryons in galaxies \citep{Bregman,Werk2014}. Anything that significantly affects the properties of the CGM plays a central role in the evolution of the central galaxy.

A number of physical processes can reduce or even remove the CGM of galaxies in dense environments: 1)  ``strangulation" or the removal of the diffuse gas reservoir of galaxies that fall into a denser environment, which then results in the eventual cessation of star formation \citep{Larson1980};  2) ``ram-pressure stripping" or the sudden and rapid removal of gas suffered by galaxies travelling at large velocities through an external gaseous medium, such as a galaxy cluster's intracluster medium or the CGM of a larger galaxy \citep{Gunn1972};  3)  ``tidal stripping" or the removal of gas due to differential gravitational forces arising from interactions with individual massive galaxies or the global host potential \citep{Merritt1983}, and 4) ``galaxy harassment" or the cumulative effects of repeated high-velocity encounters with other galaxies, which is believed to play a role in the formation of dwarf ellipticals and the destruction of low surface brightness galaxies in clusters \citep{Moore1996}. In all of these scenarios, the CGM, which lies at larger galactocentric radius than the gas or stars in the central galaxy itself, will necessarily experience a greater perturbation. 

Evidence for environmental removal of gas from galaxies is accumulating.  At a fixed stellar mass, late-type galaxies in the centers of groups lack cold H{\small I}  relative to galaxies in the group outskirts, and more massive groups show evidence of a stronger dependence of H{\small I} properties on environment \citep{Odekon2016}. The interpretation of these observations is that the H{\small I} gas is significantly stripped via ram-pressure during the galaxies' first passage through the intracluster medium \citep{Odekon2016,Jaffe2016}. Such effects are not just limited to group and cluster environments, but are also in evidence within the halo of our own galaxy \citep{Grcevich2009}. 
Again, if environmental effects can be detected in the innermost gas in galaxies, the effects on their halo gas must be catastrophic. 
To reveal the effect of environment on the CGM, we must trace the CGM in galaxies across different environments. Observing the CGM is a long-standing goal, with a correspondingly extensive literature \citep[see][for a recent comprehensive review]{CGM2017}.

The exceedingly low column density of gas in the CGM creates a significant observational challenge. Absorption line studies, which are sufficiently sensitive to sample this gas along individual sight lines, have provided the bulk of our knowledge to date on the CGM \citep[e.g.,][]{Cooksey2010,Tumlinson2011,Prochaska2011,Zhu2013,Bordoloi2014,Werk2014,Lehner2015,Prochaska2017,bordoloi}. Inferences about the population as a whole require statistical combinations of sightlines through at least tens of different galaxies \citep[e.g.,][]{Werk2014,Prochaska2017}. Unfortunately, absorption line studies cannot provide a high spatial resolution map of the CGM across a single galaxy because of the low surface density of sufficiently bright background sources. Ultimately, tracing the recombination line emission from this gas will provide such information, but our sensitivity to emission is far poorer than to absorption.

In a step toward this ultimate goal, 
we have developed a method to observe the CGM that utilizes large stacks of spectra obtained in galaxy redshift surveys \citep[][hereafter, Paper I]{zhang2016} to provide measurements of the recombination line emission from the cool ($T \sim 12,000$ K) component of the CGM. We previously presented results on the radial distribution of H$\alpha + $[N {\small II}] emission from the halos of $\sim$ L$^*$ galaxies and their neighbors to projected radii beyond 100 kpc
\citep[][hereafter, Paper II]{zhang2018} and
on the use of additional recombination lines to constrain the physical state of the gas and the ionizing mechanism  \citep[][hereafter, Paper III]{zhang2018B}.  While we are still short of our goal of mapping the CGM in individual galaxies, we have now established the magnitude of the emission luminosity and can  utilize statistical samples to constrain properties of the CGM. Emission line studies provide complementary data to absorption line studies in that the measurements have different sensitivities to density and temperature. Lastly, because our  sample consists of many thousands of galaxies, we are able to divide the same sample in various ways in our examination of the properties of the CGM.

In this work, we explore the effect of environment, if any, on the line emission from the CGM. In \S2 we discuss the two ways in which we quantify environment. Here, we also revisit simulations first presented in Paper II to connect our empirical determinations of environment to physical measures of environment. In \S3 we present significant trends in the emission line fluxes with both measures of environment, discuss whether other galaxy properties that correlate with environment could be the physical driver of these trends, present a simple, heuristic model to illustrate one scenario that explains one set of observations, discuss complicating factors in our simple interpretation, and propose how future studies could resolve the open questions our results pose.

To evaluate distances, we adopt rounded, intermediate versions of the standard $\Lambda$CDM cosmological parameters $\Omega_m$ = 0.3, $\Omega_\Lambda =$ 0.7, $\Omega_k$ = 0 and the dimensionless Hubble constant $h = $ 0.7 \citep[cf.][]{riess,Planck2018}. 

\section{Data Analysis and Selection}
We follow the 
approach developed in Papers I through III.
We obtain spectra from the Sloan Digital Sky Survey Data Releases \citep[SDSS DR12]{SDSS12} and classify galaxies that meet our criteria in redshift (now from 0.02 $\le z < $ 0.2), luminosity ($10^{9.5}\le L/L_\odot < 10^{11.0}$),  and half light radii (2 $\le {\rm R_{50}}/{\rm kpc} <$ 10) as candidate primary galaxies. 
Furthermore,  we extract measures of the galaxy's S\'ersic index ($n$)  and $r$-band absolute magnitude ($M_r$) from \citet{simard},  stellar mass, M$_*$ \citep{Kauffmann2003agn,kauffmann_mass,Gallazzi}  and star formation rates \citep{Brinchmann} from the MPA-JHU catalog. Our requirement for these ancillary measurements limits the primary sample to galaxies from SDSS DR7, but we utilize line-of-sight spectra  from DR12 to probe the CGM of these galaxies.

\subsection{Quantifying Environment}
\label{enviro}

It is always difficult to accurately quantify the environment of a galaxy. The challenge arises partly from observational limitations, such as projection effects and incomplete catalogs, and partly from our incomplete knowledge of which aspects of the environment are physically important.  
In recognition of the latter quandary, we adopt two separate approaches in estimating a galaxy's environment that mirror the long-standing debate on the relative importance of local vs. global environment \citep[e.g.][]{Dressler1980,whitmore}.

In our first approach for quantifying environment, which is aimed at quantifying the local environment, we use only the fraction of our primary galaxies that have at least one very 
massive galaxy, stellar mass $> 10^{11}$ M$_\odot$, within 2 Mpc projected separation and with $|\Delta (cz)| <$ 500 km s$^{-1}$.
The mean absolute $r$-band magnitude of these galaxies is $-22.7$, which corresponds roughly to 2.5$L^*$.
For only those primary galaxies with such a nearby massive neighbor, we quantify the environment using the projected separation between the primary galaxy and the massive galaxy, $R_{p}$. 
Using simulations to provide a greater understanding of the environment we select empirically, we find that typical galaxies selected in this manner are in environments of above-average, but not extreme, richness. 

\begin{figure}[htbp]
\begin{center}
\includegraphics[width = 0.45 \textwidth]{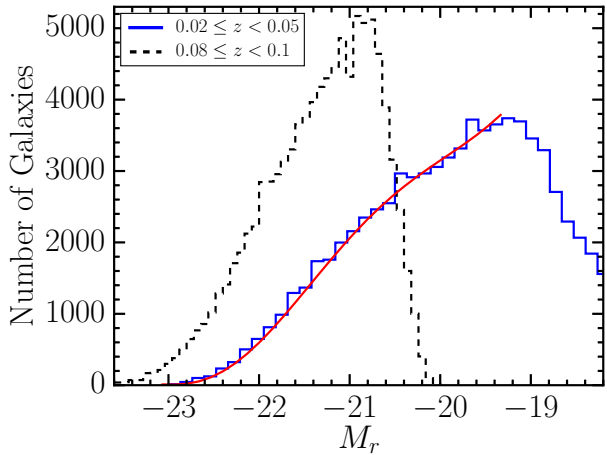} \hfill
\end{center}
\caption{$M_r$ distribution of  SDSS  galaxies with $0.02 \le z < 0.05$ and the best-fit 6th order polynomial function that we use to define the empirical luminosity function.
The luminosity function at this redshift becomes noticeably incomplete for galaxies fainter than $M_r \sim 19.5$. We use the fitted function to correct for incompleteness in our neighbor counts for primaries at larger redshifts. For comparison, the distribution of SDSS galaxies with $0.08 \le z < 0.1$ is also displayed as a dashed black line.}
\label{fig:rMag}
\end{figure}

In our second approach, which is aimed at quantifying the global environment, we define environment by counting the number of neighboring galaxies (NN) with $M_r < -19.5$, approximately 0.05$L^*$, around each primary, within 500 kpc projected separation, $r_p$, and $|\Delta (cz)| \le $ 500 km s$^{-1}$. In practice, this counting is complicated by the redshift-dependent spectroscopic magnitude limit across our volume. Specifically, the SDSS spectroscopic data are far more complete to $M_r = -19.5$ at $z = 0.02$ than at $z = 0.2$. To correct for incompleteness, we use the galaxy luminosity function determined at low redshift to estimate the number of missing galaxies down to the set limit of $M_r = -19.5$ at larger redshifts.
Rather than using the fully corrected SDSS luminosity function  \citep[eg.][]{LFBlanton,LFDorta} as our reference, 
we use the raw SDSS galaxy counts for galaxies with $0.02 \le z < 0.05$, our lowest redshift bin, which we define as complete to $M_r = -19.5$
(Figure \ref{fig:rMag}). Any incompleteness in this lowest redshift bin is irrelevant because we care about relative incompleteness as a function of redshift.  To illustrate the relative degree of incompleteness in higher redshift bins, we show the uncorrected luminosity function at $0.08 \le z < 0.1$ in the same figure. For purposes of estimating NN, we set a redshift limit of 0.1 because of high incompleteness beyond this redshift.

Our direct approach to correct for incompleteness ensures that we treat aspects that affect spectrocopic success, such as the surface brightness within the fiber and any magnitude dependence in the presence of emission lines, consistently across our redshift range. To estimate the absolute magnitude at which the survey begins to suffer significant incompleteness as a function of redshift, we divide the sample into redshift slices, measure the effective luminosity function, identify the apparent magnitude of the luminosity function turnover, and fit a 3$rd$ order polynomial function to the relation between the apparent turnover magnitude and redshift. 
For each primary, we count neighbors to the corresponding completeness limit at that redshift and integrate the empirical luminosity function to correct for the missing neighbors to $M_r = -19.5$.
For reference, the correction we end up applying is a factor of 2.2 at $z=0.08$ and 3.4 at $z=0.1$. Meanwhile, we also determine whether any neighbor has a stellar mass that is greater than that of the primary to assess whether the primary galaxy is likely to be a parent or satellite galaxy. 

\begin{figure}[htbp]
\begin{center}
\includegraphics[width = 0.48 \textwidth]{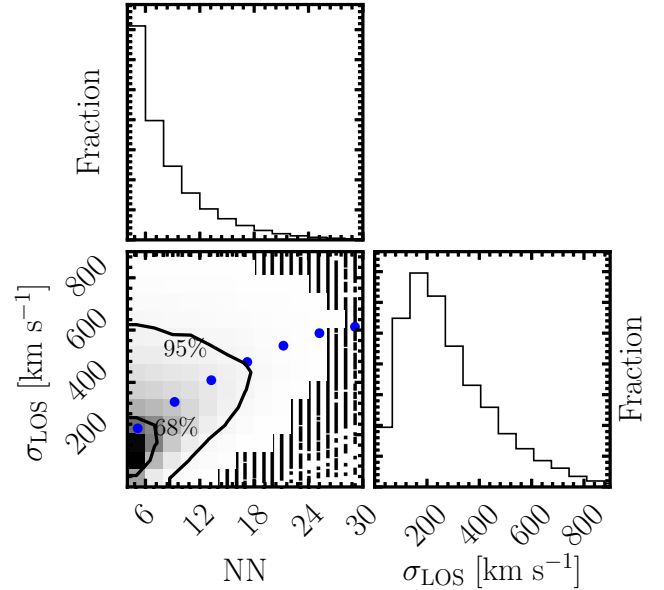} \hfill
\end{center}
\caption{Relation between NN and the line-of-sight velocity dispersion, $\sigma_{\rm LOS}$, for the environments around a primary galaxy as determined from the simulation described in the text. The blue dots represent the average $\sigma_{\rm LOS}$ vs. NN. Top and right panels show the distributions of NN and $\sigma_{\rm LOS}$.}
\label{fig:simu}
\end{figure}

To evaluate whether these counts are a valid measure of the global environment and help guide our intuition,
we assess the environment of a similarly-selected sample in the 
cosmological simulations we described briefly below. 
We identify a sample of primary galaxies that satisfy our luminosity criterion and half-light radius cut (the mean redshift of the simulated volume is arbitrary within the local universe) and then count neighbors in the manner described above, with the difference that we do not need to apply a completeness correction to the simulation. 
To estimate the relative spectroscopic completeness between SDSS and the simulation, we rank both the simulated galaxies and our SDSS primaries according to NN. We find that SDSS primaries with a certain NN correspond closely in rank to that of simulated primaries with the same NN. We conclude that the SDSS-measured NN needs no correction beyond a possible  change of $\pm$ 1 to place it on the scale of Figure \ref{fig:simu}. 

The simulations
we use, which we have used before (Paper II) and will utilize further in this paper, originate from a catalog based on halo merger trees from the {\sl Bolshoi-Planck} simulation \citep{Klypin16,RP16}, with halos found using the \textsc{Rockstar} phase-space halo finder \citep{BehrooziRockstar} and merger trees generated with the \textsc{Consistent Trees} code \citep{BehrooziTrees}, and, finally, stellar masses modeled with the \textsc{UniverseMachine} code \citep{Behroozi2018}.

The simulation offers the advantage that we can also examine the 3-D distribution of galaxies around the primaries. 
To characterize the environments of these primaries, we select all neighbors within a 2 Mpc radius sphere from the primary and evaluate the line-of-sight
 velocity dispersion ($\sigma_{\rm LOS}$) for that set of galaxies. In Figure \ref{fig:simu} we present the relationship between NN and the velocity dispersion. While there is tremendous scatter, and so for any individual environment NN is weakly constraining, NN does track environment on average. For example, the mean velocity dispersion of the environment is 350 km s$^{-1}$ for NN $= 10$,  450 km s$^{-1}$ for NN $= 15$, and 530 km s$^{-1}$ for NN $= 20$. In the same figure, we also show the numerical distribution of NN and $\sigma_{\rm LOS}$ in the simulation. Similarly, we calculate that the
 mean velocity dispersion of galaxies with $M_r < -19.5$ within a 2 Mpc radius of our defined massive galaxies is only 200 km sec$^{-1}$. This
 confirms that we are predominantly measuring the role of the local environment defined by the massive galaxy, rather than the role of a dense group or cluster environment, in our first measure of environment.
 
For both of these approaches, we examined the dependence of the results described below to somewhat different choices for the projected radius and $|\Delta (cz)|$ criteria and found no qualitative differences when these are varied by less than a factor of 2.

\subsection{Measuring CGM Emission}

To trace the CGM, we use the emission line flux of H$\alpha$ + [N {\small II}]$\lambda$6583.
We opt to only measure one of the two [N {\small II}] lines surrounding H$\alpha$ because, as we showed in Paper III, we can extract a 3$\sigma$ detection of [N {\small II}]$\lambda$6583 but not of [N {\small II}]$\lambda$6548 from the SDSS spectra. 
We measure the flux for lines of sight
at projected radii between 10 and 500 kpc from the primary.

The choice of the inner 
radius is motivated by wanting to avoid contamination from the central galaxy itself and by wanting to include as much of the CGM as possible. We have presented arguments in Paper III that 10 kpc is an acceptable choice, but we will also present results using a larger inner radius of 20 kpc.

We search the SDSS data for spectroscopic
lines of sight that satisfy the relevant $r_p$ criteria and for which the redshift of the target galaxy along that line of sight is sufficiently different from that of the primary galaxy that we avoid confusion ($|\Delta z| > $ 0.05). For each such spectrum, we fit and subtract a 10th order polynomial to a 200 \AA\ wide section surrounding the wavelength of H$\alpha$ at the primary galaxy redshift to remove the continuum flux, $C_f$. There are additional spectroscopic selection criteria, $C_f < $ 3 $\times$ $10^{-17}$ erg cm$^{-2}$ s$^{-1}$ \AA$^{-1}$, a limit on the
level of residual noise in spectrum away from the emission lines, and an emission line flux $<$  0.3 $\times$ $10^{-17}$ erg cm$^{-2}$ s$^{-1}$ \AA$^{-1}$ to eliminate contamination from interloping strong emitters such as satellite galaxies. All of these are described in more detail in Papers I-III and help reduce noise in the flux measurements\footnote{
The conversion factor to units between the values we present,  $10^{-17}$ erg cm$^{-2}$ s$^{-1}$ \AA$^{-1}$ and those used commonly in the literature to describe diffuse line emission, erg cm$^{-2}$ s$^{-1}$ arcsec$^{-2}$, is 1.7.}. 
We then sum the residual H$\alpha$ + [N{\small II}]$\lambda$6583 flux within a velocity window corresponding to $\pm 275 $km s$^{-1}$ from the primary galaxy to capture the majority of the H$\alpha$ + [N {\small II}] emission flux from the CGM gas along that line of sight. 

These measurements are then stacked for systems that satisfy specific prescribed criteria that we will discuss below. In figures and tables we present the mean emission line fluxes and the associated  uncertainties, estimated using a jackknife test. Specifically, we randomly select half of the individual spectra, calculate the mean emission line flux and repeat the process 1000 times to establish the distribution of measurements from which we quote the values corresponding to the 16.5, 50.0 and 83.5 percentiles. We compensate for using only half the sample in each measurement by dividing the $1\sigma$ estimated error by a factor of $\sqrt{2}$.

\section{Results}

Our primary goal is to measure the effect of environment on the CGM, as measured from the H$\alpha$ + [N {\small II}] emission originating in the T $\sim 10^4$K gas surrounding $\sim$ L$^*$ galaxies. We now discuss the results from each of the two characterizations of environment.

\subsection{Dependence on $R_P$}
\label{local}

\begin{figure}[htbp]
\begin{center}
\includegraphics[width = 0.48 \textwidth]{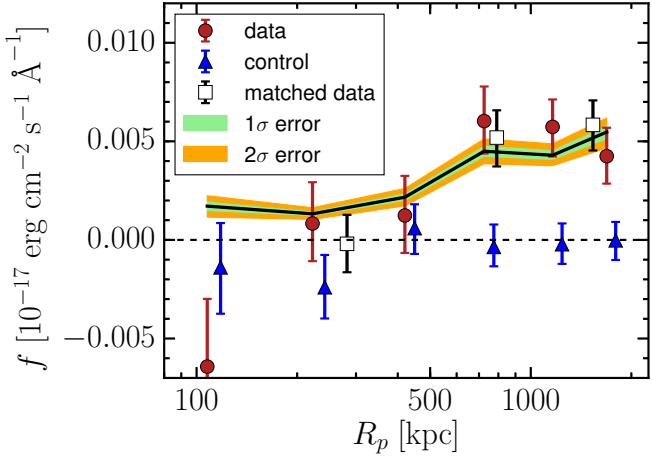} \hfill
\end{center}
\caption{H$\alpha$+[N II] flux for $10 < r_p/{\rm kpc} < 50$ as a function of separation $R_p$ from a massive galaxy for the complete galaxy sample (red circles)  and for a stellar mass and morphology matched galaxy sample (black open squares).
These measurements are to be compared to the fluxes measured over fictitious primaries at the same $R_p$ from the same set of massive galaxy (blue triangles). For easier visualization, we apply slight horizontal offsets to the control sample. The latter are, as expected, consistent with zero net flux. The shaded region corresponds to 1$\sigma$ and 2$\sigma$ uncertainty of the model prediction described in \S\ref{model}. Primaries at large separations from the massive neighbor have detectable emission line flux at small radii, while both the control and primaries within 500 kpc of a massive galaxy do not.}
\label{fig:sep_app2}
\end{figure}

In this characterization of environment, we consider only those primaries within a projected distance of 2 Mpc and a  recessional velocity difference of 500 km s$^{-1}$ from a massive ($M_* > 10^{11}$ M$_\odot$) galaxy. We measure the H$\alpha$ + [N {\small II}] emission line flux for $10 < r_p \le 50$ kpc and examine that quantity as a function of the projected distance from the primary to the central massive galaxy, $R_p$, in Figure \ref{fig:sep_app2}.  We find  distinctly lower emission line flux in the inner $R_p$ bins, with a larger and nearly constant flux level for $R_p > 500$ kpc. The one surprising aspect of these data is the slightly significant ($\lesssim 2 \sigma$) negative flux values for the innermost data point, suggesting mean absorption rather than emission at H$\alpha$. This result could in part be due to the stellar halos of the massive galaxies, but the result is statistically marginal and critically dependent on the difficult measurement of the baseline continuum level, as we mentioned earlier in a different context.

\begin{deluxetable}{cc}
\tablewidth{0pt}
\tablecaption{The H$\alpha$ + [N {\small II}] emission flux vs. $R_p$}
\tablehead{  \colhead{$R_p$ [kpc] \tablenotemark{a}} &  \colhead{FLUX [$10^{-17}$ erg cm$^{-2}$ s$^{-1}$ \AA$^{-1}$]\tablenotemark{b}}
  }
\startdata
108 & ($-6.42 \pm 3.28)\times 10^{-3} $\\
222 & ($0.83 \pm 1.94)\times 10^{-3}  $\\
420 & ($1.24 \pm 1.94)\times 10^{-3}  $ \\
1162 & ($5.73 \pm 1.51)\times 10^{-3}  $ \\
1687 & ($4.25 \pm 1.23)\times 10^{-3} $
 \enddata
\label{tab:approach2}
\tablenotetext{a}{Distance between primary and massive, $M_* > 10^{11}$ M$_\odot$, neighbor.}
\tablenotetext{b}{Integrated flux for $10 < r_p /{\rm kpc} \le 50$.}
\end{deluxetable}

To understand the measurements better and explore the possibility of contamination from the massive galaxy, we construct a control sample in which we insert two artificial primaries at empty positions, but at the same $R_p$, around each massive galaxy. We search for H$\alpha$+[N {\small II}] emission at the redshift of the associated actual primary and stack the results. In Figure \ref{fig:sep_app2}, we have also included the measured emission line fluxes for these false primaries. As expected, if there is no contamination from the massive galaxy, we find values consistent with zero flux and no dependence with $R_p$. We confirm that the steep decline in flux toward small $R_p$ is associated with changes in the CGM of the primary, although the slight systematic tendency for the control to produce negative fluxes suggest that there may be a modest but systematic overestimate of the continuum.

We examine the stellar mass and S\'ersic index distribution for the primaries at different $R_p$ to
determine whether other factors could give rise to the observed result. 
 We find that the stellar mass and S\'ersic index distribution of the primaries are nearly independent of $R_p$ (indistinguishable to our level of resolution). To test this conclusion, we construct a stellar mass and morphology matched samples. To do
this, we force the percentage of systems for each stellar mass and S\'ersic index bin to be the same in each $R_p$ bin. This requirement ensures  that  the  samples  are  mass and
morphology balanced, although at the cost of lower signal to noise ratio.
To improve the statistics, we combine the three inner radial bins into one and split the outer three radial bins into two. The matched-sample results are shown in Figure \ref{fig:sep_app2} and are indistinguishable from the original, confirming that the internal properties of the primary galaxies are not driving the trend seen in Figure \ref{fig:sep_app2}. 
 We do not interpret this result to be in conflict with previous measurements of a correlation between morphology and $R_p$ \citep[e.g.,][]{Wetzel2012},  but instead attribute the difference in results to our limited range in $R_p/r_{\rm vir}$ and small sample size.

\subsubsection{A Simple Model}
\label{model}

In Paper II, we exploited a straightforward model to better understand the origin of the emission line flux and estimate the cool gas fraction. Here, we employ the same basic model with some modifications to explore a scenario for the result presented in Figure \ref{fig:sep_app2}. The model serves as an order-of-magnitude calculation. It acts as a rough check on  our understanding of the situation, can help guide intuition, and provides a framework with which to formulate the next set of inquiries. It is not a replacement for a detailed, physics-driven model.

Our basic premise is that the CGM from 10 to 50 kpc is affected, possibly by ram-pressure, as the galaxy ventures closer to the massive neighbor. We heuristically model that as an inner region around the massive galaxy in which satellite galaxies have depressed CGM emission out to a characteristic physical separation, $R_C$, beyond which the cool gas fraction is constant. In this model, as in Paper II, we adopt the cool gas temperature to be 12,000 K, but discuss the effect of temperature variations further below, and adopt a gas density profile about the target galaxy consistent with that describing gas in an NFW potential.
Specifically, we model the cool gas fraction as linearly rising from 0 at a physical radius, $R$, equal to 0 to a value of $a$ at $R = 500$ kpc, and continuing to rise if $R_C > 500$ kpc. 
At $R_C$ the cool fraction converts to the asymptotic value, $c$. We do not require continuity in $C_f$ at $R_C$ but we do require that $C_f$ not exceed $c$ at any $R$. This is expressed mathematically as  follows:
\begin{equation}
C_f(R) = 
  \begin{cases}
    a({R\over{500 \ {\rm kpc}}}) & \quad \text{if} ~R < R_C\\
    c &  \quad  \text{if} ~R \ge R_C
  \end{cases}
\end{equation}

Once we define the priors for the parameters, we perform a Bayesian analysis to derive confidence intervals on each of the three parameters.
Based on the estimation of the cool fraction in Paper II, the parameter $c$ is expected to be $\sim$ 0.36. 
We adopt uniform priors for the three parameters for simplicity as follows:
\begin{equation}
400 < R_C < 1100, \quad 0 < a < 0.4, \quad 0.2 < c < 0.6
\label{eq:prior}
\end{equation}
 The posterior distribution, $p(\Theta|{\rm data})$, for the parameters based on the data is described as follows:
\begin{equation}
p(\Theta|{\rm data}) = \frac{p(\Theta) \cdot \Pi ~p({\rm data}|\Theta)}{p({\rm data})}
\end{equation}
where $\Theta$ is the parameter space $\Theta = (R_C, a, c)$, $p(\Theta)$ is the prior distribution described in Eq. \ref{eq:prior},  $p({\rm data}|\Theta)$ is the likelihood of the data and $p(\rm data)$ is the marginal probability for the data. 

We define the likelihood of obtaining the data given a specific model using the difference between the actual and model data, ${\rm exp[-(actual - model)^2/\sigma^2}]$, where $\sigma$ is the observational uncertainty. We use a package called ``{\it emcee}" \cite[]{emcee}, which implements a Markov chain Monte Carlo sampling of the likelihoods across parameter space to calculate the posterior distribution. {\it emcee} is a {\it Python} implementation of the affine-invariant ensemble sampling approach suggested by \cite{Goodman2010MCMC}. It utilizes an ensemble of N  walkers, and it will evolve each for a certain number of steps.  We initialize each walker (total 500 walkers in our simulation) by randomly sampling from our prior distributions, and evolve each walker for 500 steps. We discard the first 100 steps since it takes a certain number of steps ($\sim 50$ in this study) for the results to become stable. 

\begin{figure}[htbp]
\begin{center}
\includegraphics[width = 0.48 \textwidth]{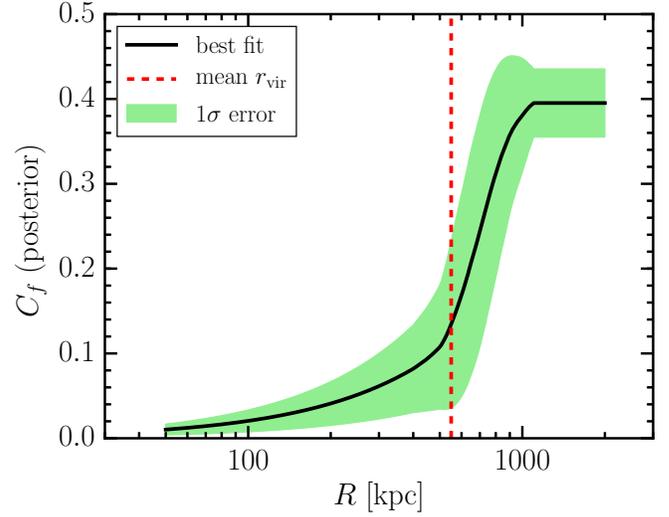} \hfill
\end{center}
\caption{Posterior distribution for cool gas fraction in the CGM of the target galaxy, within 10 and 50 kpc, as a function of distance, $R$, from the massive neighbor. The shaded region highlights the 1$\sigma$ uncertainties. The red vertical dashed line represents the mean virial radius for the massive neighbors.}
\label{fig:Cf}
\end{figure}

\begin{figure*}[htbp]
\begin{center}
\includegraphics[width = 0.675 \textwidth]{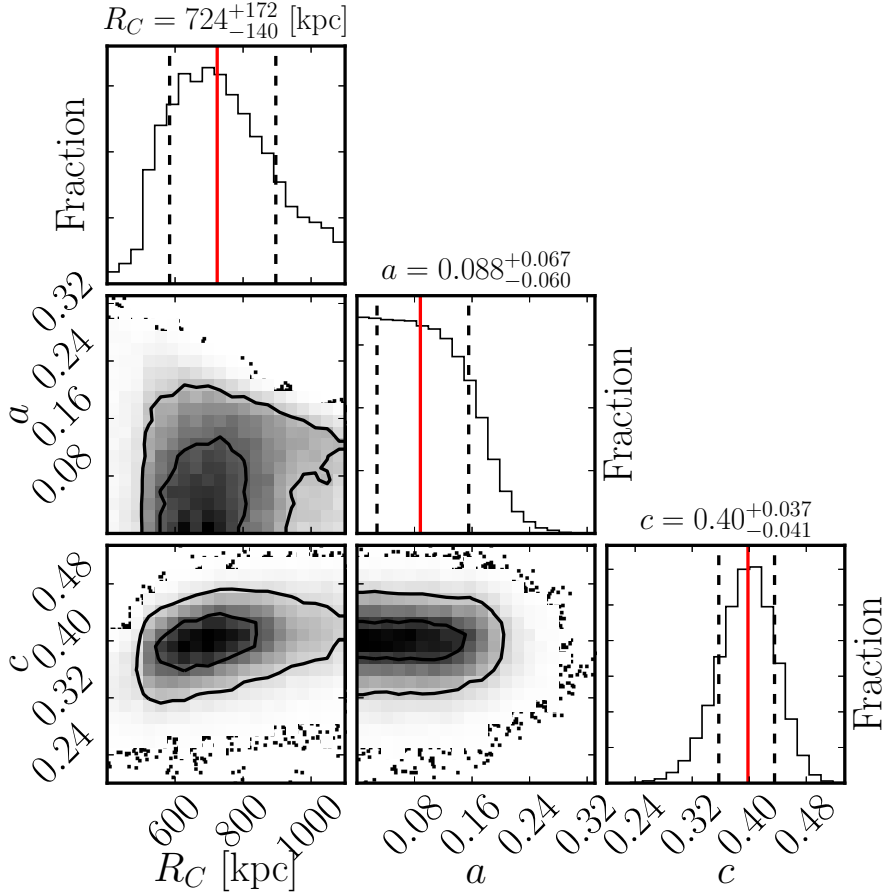} \hfill
\end{center}
\caption{Posterior probability distributions for model parameters $R_C$, $a$ and $c$, and the correlations between them. We display the median value (red vertical line),  lower  error estimated by the 16\% value and higher error estimated by the 84\% value (dashed black vertical line) for the model parameters $R_C$, $a$ and $c$. The two contours in the correlation plot represent 68\% (small one) and 95\% (big one) confidence levels. }
\label{fig:model}
\end{figure*}

Figure \ref{fig:model} displays the posterior distributions of the three parameters, as well as the median value and the 1$\sigma$ confidence level (based on the 16th and 84th percentiles),  and the correlations between them.  The preferred value of $R_C$, 724$^{+172}_{-140}$ kpc, indicates that the CGM of primary galaxies within $\sim$ 700 kpc of a massive neighbor is significantly effected.
The preferred value of $c$, $0.40^{+0.037}_{-0.041}$, is consistent with what we had measured previously (Paper II) as the cool fraction for the entire sample.
The preferred value of $a$, 0.088$^{+0.067}_{-0.060}$, where $a$ is defined as the cool fraction at $R = $ 500 kpc, indicates a severe depletion of cool gas even out to a distance of 500 kpc from the massive neighbor. The virial radii of the massive galaxies range from $\sim$ 400 kpc to $\sim$ 1000 kpc, with an average of 550 kpc.

We interpret these results as indicative of the removal of the CGM of the target galaxy by the massive neighbor.
We show the posterior distribution for the cool gas fraction in the CGM of the target galaxy, between 10 and 50 kpc, as a function of distance from the massive neighbor in Figure \ref{fig:Cf}. The effect is most  severe when the target galaxy lies within the virial radius of the massive neighbor,  but begins to be noticeable outside of this radius.  The model suggests that satellites rapidly lose their gaseous haloes, or that their hot gaseous halos quickly become unable to cool after reaching the virial radius of a larger halo. The apparent onset of the effect at radii larger than the virial radius could reflect the nature of the splashback radius \citep[for discussion see][]{more}, where galaxies that have fallen inside of the virial radius return to larger radii as they execute their orbits, or on the affected nature of mass accretion onto halos at these radii \citep{Behroozi2014}.

\subsection{Dependence on NN}

To explore the dependence of CGM emission on global environment, we split the data into three categories according to the number of neighbors (NN). In Figure \ref{fig:simu} we showed the correlation between NN and velocity dispersion in the simulations and in \S\ref{enviro} we discussed how to connect this to our SDSS measurements of NN.  From those, we know that NN $< 5$ corresponds roughly to isolated galaxies with environmental velocity dispersion less than 200 km s$^{-1}$, 5 $\le$ NN $<12$ corresponds to poor groups with 200 $<\sigma <$ 400 km s$^{-1}$, and NN $\ge 12$ corresponds to rich groups and clusters with $\sigma >$ 400 km s$^{-1}$. We split our sample into these three categories with 1,085,740 in the first, 166,214 in the second, and  16,762 in the third. 
The qualitative nature of the results is not sensitive in detail to slight changes in the NN ranges.

Before calculating the fluxes in these different environment categories, we consider various subtle dependencies that could affect our results. First, correlations between environment and a galaxy's mass or morphology may play a role. If galaxies in richer environments tend to have a less pronounced CGM for other reasons that correlate with environment, it may appear to be the case that environment is driving the lack of a CGM. For example, if early type galaxies, which are more common in dense environments, have a less pronounced CGM for due to their formation history, then that could create results that appear to indicate an environmental effect. In practice, absorption line studies of the CGM of luminous red galaxies, $L \ge 3L^*$ find a significant cool component, comparable to what is found in L$^*$ star-forming galaxies \citep[e.g.,][]{Gauthier2009,Zhu2014,chen2018,Zahedy2018}, but we should nevertheless control for mass and morphology because they are known to correlate with environment.

\begin{figure}[htbp]
\begin{center}
\includegraphics[width = 0.45 \textwidth]{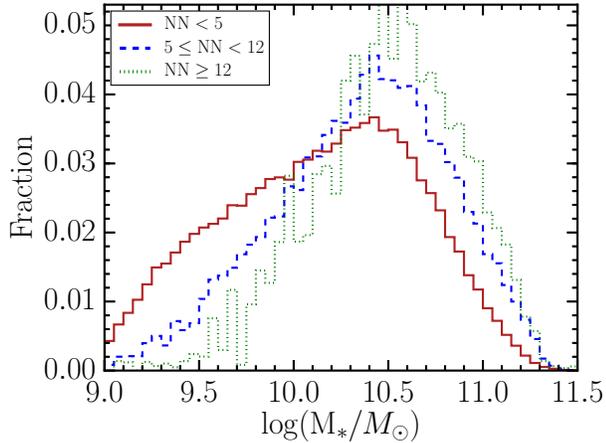} \hfill
\end{center}
\caption{Distribution of primary galaxy stellar mass as a function of NN.
As anticipated, galaxies in denser environments tend to have larger stellar masses than those in lower density environments.}
\label{fig:sm}
\end{figure}

\begin{figure}[htbp]
\begin{center}
\includegraphics[width = 0.45 \textwidth]{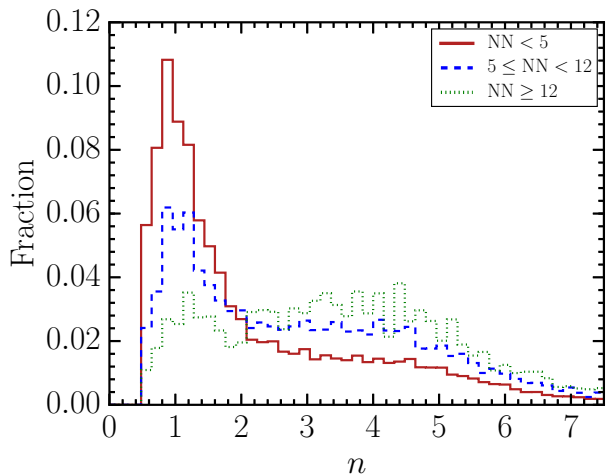} \hfill
\end{center}
\caption{Primary galaxy S\'ersic index distribution as a function of NN.   Galaxies in denser environments are more heavily weighted toward early types than those in lower density environments.}
\label{fig:sersic}
\end{figure}

To emphasize this point, 
we compare the stellar mass and S\'ersic index distributions for the primary galaxies in the three environmental categories (Figures \ref{fig:sm} and \ref{fig:sersic}, respectively). Indeed, as has been well-documented, there are systematic differences in the masses and morphologies of galaxies as a function of environment \citep[e.g.,][]{Roberts1994,Conselice2014}. 
To determine if the differences we see in CGM profiles are driven by the mass or the morphology differences, we construct
subsamples of primaries with the same stellar mass and morphology distribution. We do the stellar mass and S\'ersic index matching for each of the three environment categories. As we did before in a different context, we force the percentage of the systems for each stellar mass and S\'ersic index bin to be the same for the three categories. 

\begin{figure}[htbp]
\begin{center}
\includegraphics[width = 0.45 \textwidth]{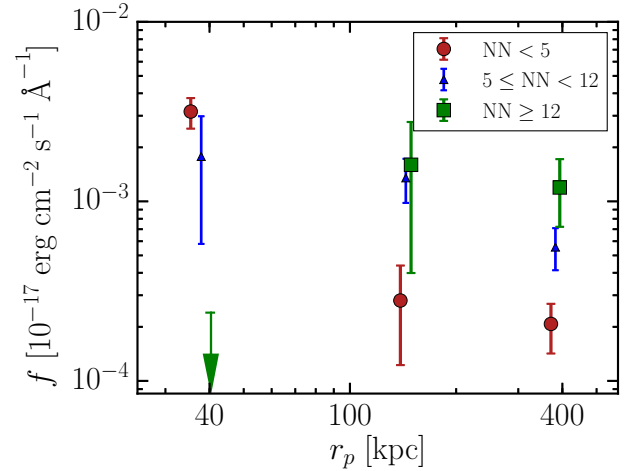} \hfill
\end{center}
\caption{H$\alpha$ + [N {II}]$\lambda$6583 flux as a function of $r_p$ for the three environmental categories based on number of nearby neighbors, NN, that are now stellar mass and morphology matched. The measurement for the flux in the innermost bin for primaries in the richest environments is consistent with zero and shown as an upper limit. For easier visualization, we apply slight horizontal offsets to the different samples. The errorbar for the flux in the inner radius bin for the NN $\ge 12$ subsample has been inflated relative to what is in Table \ref{tab:approach1} to account for an estimated systematic uncertainty in the determination of the background (see Paper II for details).}
\label{fig:app1_match}
\end{figure}

We present the H$\alpha$+[N {\small II}] flux vs. $r_p$ for the three environmental categories described above in Table \ref{tab:approach1} and in Figure \ref{fig:app1_match}.
The broadest result visible in the figure is that the emission line flux behavior interior and exterior to $r_p \sim 50$ kpc as a function of environment reverses. In 
our innermost $r_p$ bin, $10 < r_p \le 50$ kpc, the flux drops when
we consider primaries in the richest environment. 
For projected radii between $\sim$ 100 and 500 kpc, we find the opposite behavior in that the emission flux is larger for galaxies in denser environments. We discuss the statistical significance of the behavior at $r_p < 50$ kpc below, when we present these results in a slightly different way.

\begin{deluxetable*}{cccc}
\tablewidth{0pt}
\tablecaption{H$\alpha$ + [N \small{II}] fluxes for mass and morphology matched samples}
\tablehead{
\colhead{NN\tablenotemark{a}} &  \multicolumn{3}{c}{FLUX [$10^{-17}$ erg cm$^{-2}$ s$^{-1}$ \AA$^{-1}$]} \\
& \colhead{$10 < r_p/{\rm kpc} \le 50$}  & \colhead{$50 < r_p/{\rm kpc} \le 200$}  & \colhead{$200 < r_p/{\rm kpc} \le 500$}   }
\startdata
$< 5$ & ($3.19 \pm 0.56)\times 10^{-3}$ & ($0.28 \pm 0.16)\times 10^{-3}$ & $(0.21 \pm 0.06)\times 10^{-3}$  \\
 $5-11$ &  $(2.25 \pm 1.44)\times 10^{-3}$ &  $(1.35 \pm 0.37)\times 10^{-3}$ & $(0.56 \pm 0.15)\times 10^{-3}$ \\
  $\ge$12 &  $(-6.82\pm 4.45)\times 10^{-3}$ &  $(1.60 \pm 1.18)\times 10^{-3}$ & $(1.20 \pm 0.44)\times 10^{-3}$ 
\enddata
\label{tab:approach1}
\tablenotetext{a}{NN stands for number of neighbors.}
\end{deluxetable*}

Our initial interpretation of these results, in combination with the results of Paper II, which demonstrated that the emission line flux beyond 50 kpc is in general dominated by the CGM of nearby galaxies, is that emission at small radii is depressed in the denser environments, presumably due to the lack of cool gas  but perhaps also due to other factors  that affect the emission flux such as gas temperature  or local density differences, and that the emission at large radii is enhanced in the denser environments, presumably due to increased contamination due to a relative excess of nearby galaxies. We note that this situation complicates the general interpretation of results for the CGM at large projected separations. For example, \cite{Johnson2015} found a slight increase in the O VI covering fraction at large radii in denser environments, which is qualitatively consistent with our finding, but reached different conclusions regarding the origin of the effect. Because of this added complexity, we will focus on the measurements out to 50 kpc to explore the effect of environment on the CGM of galaxies, rather than on the possibility of flux contributions by neighbors.

\begin{figure*}[htbp]
\begin{center}
\includegraphics[width = 0.76 \textwidth]{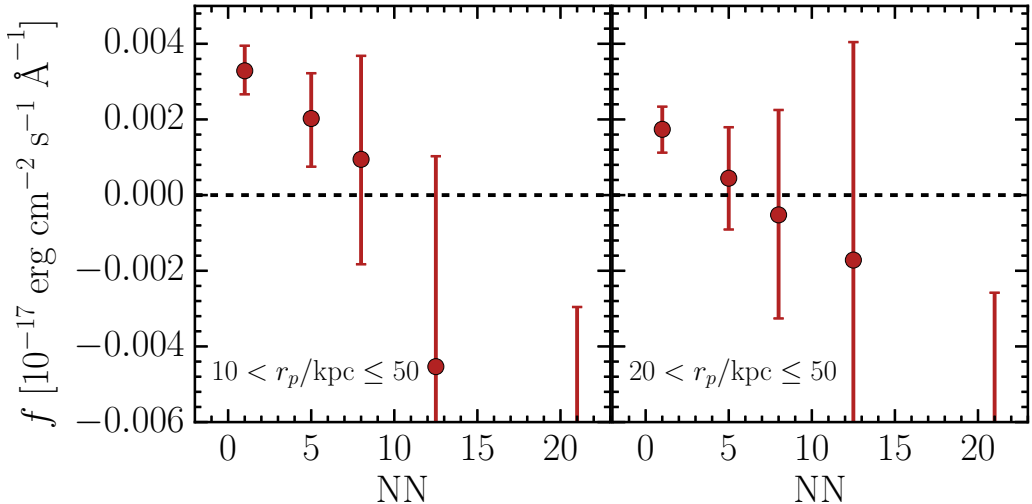} 
\end{center}
\caption{The integrated flux of H$\alpha$ + [N {\small II}]$\lambda$6583 between 10 and 50 kpc (left panel) and between 20 and 50 kpc (right panel) as a function of NN for the mass and morphology matched sample.  The uncertainty in the overall normalization, which depends on a precise measurement of the contamination and background flux level, is not captured in the error bars, which is why some measurements are negative.}
\label{fig:app1_NNmatch_20kpc}
\end{figure*}

We now return to the question of whether the evidence for depressed flux at small radius in denser environments is statistically significant. The simple consideration of how often a random draw from the larger sample (NN $<$5 bin) results in a result that is at least as different as that found for smallest sample (NN $\ge$ 12 bin) suggests that the difference between the measurements for these two bins is significant at the 95\% confidence level. However, this test neglects the data at intermediate environments, so we also do the following test. We divide the data more finely in terms of NN, still performing the stellar mass and morphology matching, and present those measurements in the left panel of Figure \ref{fig:app1_NNmatch_20kpc}. The flux consistently decreases with increasing NN. The one surprising aspect of these data is the negative flux values at large values of NN. These measurements are consistent with complete gas removal, zero flux, at the 2$\sigma$ level, but perhaps also indicative of the difficultly in precisely measuring the baseline continuum level and the determination of the background.

As we stated before, one concern in interpreting our result is that we might have mixed measurements from different radial regimes, the innermost of which may be contaminated by emission from the central galaxy and unrelated to the CGM. So, we repeat the measurements using an inner radius cutoff of 20 kpc (right panel in Figure \ref{fig:app1_NNmatch_20kpc}). 
From a rank correlation test of the data presented in Figure \ref{fig:app1_NNmatch_20kpc}, we find 
that the trend did not arise randomly with greater than 99.99\% confidence for either an inner $r_p$ cutoff of 10 or 20 kpc.
We conclude that we find a statistically significant relation between decreasing flux from the cool CGM component within 50 kpc and increasing environmental richness and that this result is not due to correlations between galaxy stellar mass or morphology and environment.

We close this topic by noting that our earlier model (\S\ref{model}) is incomplete. It fails to reproduce the results relating flux to NN. Because only a small fraction of primaries are near a massive neighbor, this model predicts that most galaxies are independent of larger-scale environmental effects. As such, it is unsurprising that it fails to reproduce the NN results. We interpret the failure of the model to mean that a galaxy's CGM is affected both by its local environment (i.e. whether it is a satellite of a massive neighbor) and by its global environment (i.e. whether it lies in a generally rich environment). The latter was not included in our simple model. More complete and sophisticated models that simultaneously address a large number of empirical results \citep[e.g.,][]{Xie2018} should also 
aim to reproduce the results presented here.

\subsection{Complications}

There are a number of complicating factors in our measurement of the CGM and the interpretation of our empirical results. 

\subsubsection{H$\alpha$ and [N {\small II}]}

We have shown in Paper III that
the ratio of H$\alpha$ to [N {\small II}] changes as a function of primary galaxy mass. We ascribed that variation to changes in the state of the gas due to differences in the ionizing mechanism. Here we have measured the sum of those two lines. There is no expectation that an equivalent amount of CGM will yield a similar summed flux if the ionization sources are changing. Given that we expect changes in the mean masses of our primaries as a function of environment, this presents a challenge to our measurement of the CGM and the interpretation of our results. 

We have attempted to address this issue by examining mass-matched samples. However, it is possible that the variation in ionizing sources is a combination of mass and some other factor, including, possibly, environment. If so, then our observation may not be entirely reflecting differences in the CGM cool gas fraction.

\begin{figure*}[htbp]
\begin{center}
\includegraphics[width = 0.76\textwidth]{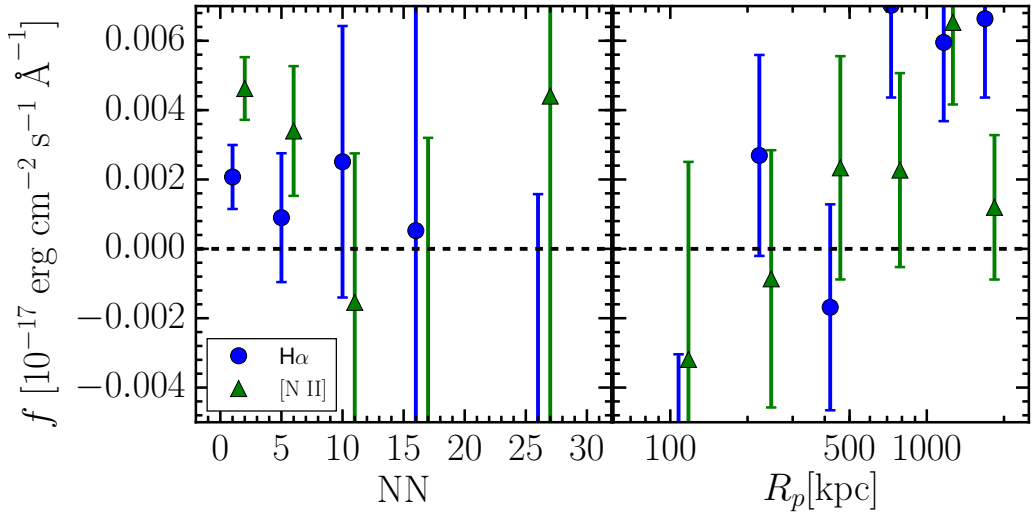} \hfill
\end{center}
\caption{The integrated flux of H$\alpha$ (blue circle) and [N {\small II}] $\lambda$6583 (green triangle) between 10 and 50 kpc as a function of NN (left panel), and the integrated flux of H$\alpha$ (blue circle) and [N II] $\lambda6583$ (green triangle) within $10 < r_p/{\rm kpc} < 50$ as a function of separation $R_p$ from a massive galaxy (right panel). Although the results are noisier for the individual lines, the trends they show are consistent with each other. For easier visualization, we apply slight horizontal offsets to the different samples.}
\label{fig:Ha_NII}
\end{figure*}

Separately, H$\alpha$ and [N {\small II}] show the same behavior with environment as H$\alpha$ + [N {\small II}], although the results are noisier (Figure \ref{fig:Ha_NII}).  The consistent results when using H$\alpha$ and [N {\small II}] individually, help confirm the results obtained using the combination. 

\subsubsection{Temperature, Density, and Gas Mass}

Another potential complication in interpreting the results is that recombination line fluxes depend on total gas mass, gas density, and temperature. Modifications in the cool gas fraction imply only a change in total gas content. H$\alpha$ recombination is suppressed as the temperature rises. The measured fluxes in the $10 < r_p < 50$ kpc bin differ by a factor of 5 between isolated primaries and those in the densest environment (in the mass and morphology matched sample). Because of the nearly linear dependence of the recombination rate on gas temperature at this temperature, such a flux difference would imply a factor of $\sim$ 5 difference in temperature, if all other parameters are equal. 
 This would appear to be an unreasonably large temperature change to expect from a process that would not affect the gas mass or density. We conclude that the decrease in flux could perhaps arise from a change to a combination of gas content, density, and temperature, rather than solely to a change in gas content, but that it is unlikely that the change is entirely driven by changes in the gas temperature.

\subsubsection{Clumpiness}

Changes in local gas density are potentially more problematic as the emission depends on density squared.
In the modeling we present, we assume a smooth distribution of cool gas. We only change the fraction of gas in this component, $C_f$. However, an additional degree of freedom that should be considered is the clumpiness of the cool gas component, which would change the local density. For a given total mass of cool gas, one can increase the local density of this gas by clumping it. The recombination rate, and therefore the emission line flux, increases as density squared, and so clumpiness would result in lowering the empirical constraint on $C_f$. In such a scenario, some lines of sight might entirely miss the cool component and have no flux, while others would have elevated flux (up to our imposed limit of 0.3 $\times$ $10^{-17}$ erg cm$^{-2}$ s$^{-1}$ \AA$^{-1}$), relative to that in the smooth models. As an example, consider what happens if all of the gas within $r_p$ is clumped into a sphere of radius 0.5$r_p$. The CGM coverage fraction out to $r_p$ would decrease by a factor of 4, so only 1 in 4 lines of sight would intersect the cool component. However, the density would rise by a factor of 8. The recombination rate increases as density squared, so the emission flux observed in lines of sight that intersect the sphere would be 64 times larger than previously. Taking this increase and the smaller covering factor into account, the mean flux obtained over many lines of sight would be 16 times higher than in the smooth model. Therefore, the sample would have a higher stacked flux if clumpiness is increased, for a given fixed $C_f$.  A simple test for clumpiness would be to measure the variance in emission line fluxes among lines of sight. Unfortunately, with the current data we can only measure the mean fluxes in stacked samples with sufficient precision. 
An independent constraint would come from absorption line studies, which are sensitive to covering fraction and column density but not local density.

Absorption line studies provide many measurements of the covering fraction. For example, 
\cite{Chen2010} find that $\sim$80\% of ordinary galaxies possess Mg II 2796\AA\ absorbers with equivalent width $\ge 0.1$ \AA ~ and \cite{Johnson2015} measure
a mean covering fraction of  0.89 for H {\small I} absorption within the virial radius of faint galaxies ($L > 0.1L^*$). 
Both results, as well as others in the literature, suggest that the CGM has a fairly high covering fraction. Such results are consistent, to the degree that they are consistent with a covering fraction of one, with our simple smooth model for the CGM. While they do hint at some clumpiness, covering fraction and clumpiness are not easily related. The CGM could have large local density fluctuations and still provide full coverage. On the other hand, deviations from a covering fraction of one do indicate that the CGM has some structure and so suggest that $C_f$ is likely to be somewhat smaller than we infer from our simple model.

Variations in clumping could be used to explain, at least in part, our observations.
The lower emission line fluxes found in denser environments could result from the cool gas being less clumpy in the halos of such primaries. Perhaps interactions, either gravitational or hydrodynamical, have shredded the cool gas clouds and distributed the gas more smoothly within the halos (although it is difficult to imagine that the shredded gas would remain cold). The effect of interactions could be quite complex, as evidenced by the complete lack of O IV absorption at small impact parameter in one set of interacting galaxies \citep{johnson14}.

\subsection{Using Absoprtion and Emission Line Studies}

While absorption line studies will certainly help inform our interpretation of the emission line fluxes, the converse may also prove valuable. For example, 
the 
 non-detection of absorption line systems associated with an individual galaxy that lies a few hundred kpc from a massive neighbor, which might have previously been interpreted in terms of the intrinsic CGM properties, can now be attributed to the environmentally-caused reduction of the CGM of this galaxy.
 Observational expense and a lack of suitable background sources have generally 
 limited the sample size of absorption line studies to $\lesssim 100$ lines of sight, making it difficult to simultaneously control for a variety of parameters such as galaxy mass and environment. An exception to this state of affairs are the SDSS stacked absorption studies \citep{Zhu2013,Zhu2014}, but the most detailed constraints come from small samples of high resolution spectra. Emission line studies can help
 establish the broad dependence of CGM properties with parameters such as mass and environment that can then be used to help interpret detailed absorption line results. 
 
The two types of diagnostics have complementary strengths. Emission line studies, such as that done here, can easily provide a test for a hypothesized dependence. In Papers II through this one, we have been able to examine the role of neighbors, mass, and environment
on the CGM properties without requiring a single additional observation. One could easily revisit the data in a search for other speculated behavior. In contrast, a strength of absorption line studies is that they are so sensitive that results can be obtained for individual galaxies, allowing one to measure galaxy-to-galaxy differences. For example, variance among CGM properties as determined from absorption line results for isolated galaxies can help constrain the clumpiness of the CGM. The future of CGM studies relies on combining the results of limited, but much more detailed, absorption line studies, with the global population studies that can be done with emission lines. Ultimately, emission line studies of individual galaxies will be possible, but such observations are most likely to be the domain of the next generation of ground based telescopes.

Some overlap in results is starting to occur.
\cite{Burchett2016} reported that $\sim 60\%$ of galaxies with $z < 0.055$ in low-density regions (defined to be regions with fewer than seven $L > 0.15 L^*$ galaxies within 1.5 Mpc) have affiliated C IV absorption at projected radii $<$ 160 kpc, while none (0/7) of the galaxies in denser groups with halo mass greater than $10^{12.5} M_\odot$ do. The sample and sightline selection do not match directly to those we have described, but both absorption and emission tracers are showing evidence of a diminished cool CGM component in high density environments. 
A detailed joint study is beyond this paper, but a quantitative comparison of the emission and absorption line tracers would address the question of density and temperature differences because the two phenomenon depend differently on those parameters. 
\section{Conclusions}

We present a study of the dependence of the CGM on the environment that a galaxy inhabits using stacked emission line measurements. 
We use two methods to characterize the environments: 1) the separation between the primary (the galaxy whose CGM we are probing) and its parent galaxy, where the parent is defined to be massive galaxy with stellar mass greater than $10^{11}$ $M_\odot$ and 2) the number of neighboring galaxies. We utilize cosmological simulations to assess how our empirical environment markers track environment. 
In the first approach, we are primarily probing the effect of isolated parent galaxies on the CGM of their satellite galaxies. In the second approach, we are primarily probing the effect of the global environment, because the number of neighbors correlates, with large scatter, with the velocity dispersion of the environment. 

We find that the H$\alpha$ + [N {\small II}] emission flux in the innermost radial bin, $10 < r_p < 50$ kpc, which we expect to be relatively free from contamination from nearby galaxies (Paper II), drops sharply both when we consider primaries closer to their massive parent galaxy and primaries in richer environments. These results show a statistically significant effect of environment on the CGM properties. This result is reproduced if the innermost radial bin is defined to be $20 < r_p < 50$ kpc.
We confirm that this result is not driven by other galaxy properties, mass or morphology, that correlate with environment.

A plausible, but not unique, interpretation of this distinctly lower emission line flux from the CGM of Milky-Way-like galaxies in rich environment is that the CGM has been removed. 
To explore this hypothesis, we exploited the straightforward model developed initially in Paper II. We assume a linear relation between the  remaining cool gas and distance between primary and parent. We run an MCMC model to constrain the model parameters and find parameters that reproduce the observations of emission line flux with distance from the parent. In this model, the cool gas component is significantly reduced within a distance from the massive parent, $M_* > 10^{11} M_\odot,$ of $\sim$ 700 kpc, with about 80\% of it being removed even within 500 kpc. This model fails to reproduce the behavior observed in the CGM emission line flux with the number of neighbors. We suggest that there is an additional effect related to the global environment that was not included in our simple model. The lack of emission not only indicates the lack of cool gas, but also the inability of any hotter gas that might be present to cool. As such, the pathway to refueling any subsequent star formation appears stalled.

As we stated above, there are a number of alternate explanations for a decline in emission line flux
in certain environments, such as systematic differences in density fluctuations due to clumpiness or temperature differences in the CGM. With stacked measurements, which obscure system-to-system differences, it is difficult to resolve these concerns. However, in combination with absorption line studies, which have different sensitivities to density and temperature fluctuations, there is a path forward. The stacked emission line studies allow large parameter studies, such as that with environment described here, while the much smaller, but more detailed, absorption line studies address questions regarding the detailed physics of the CGM. Together, the two provide a new opportunity to 
explore the critical CGM of galaxies.

\section{Acknowledgments} 
DZ and HZ acknowledge financial support from 
 NSF grant AST-171384. JW acknowledges support from a 2018 Sloan Foundation Fellowship. The authors gratefully acknowledge the  SDSS III team for providing a valuable resource to the community.

Funding for SDSS-III has been provided by the Alfred P. Sloan Foundation, the Participating Institutions, the National Science Foundation, and the U.S. Department of Energy Office of Science. The SDSS-III web site is http://www.sdss3.org/.

SDSS-III is managed by the Astrophysical Research Consortium for the Participating Institutions of the SDSS-III Collaboration including the University of Arizona, the Brazilian Participation Group, Brookhaven National Laboratory, Carnegie Mellon University, University of Florida, the French Participation Group, the German Participation Group, Harvard University, the Instituto de Astrofisica de Canarias, the Michigan State/Notre Dame/JINA Participation Group, Johns Hopkins University, Lawrence Berkeley National Laboratory, Max Planck Institute for Astrophysics, Max Planck Institute for Extraterrestrial Physics, New Mexico State University, New York University, Ohio State University, Pennsylvania State University, University of Portsmouth, Princeton University, the Spanish Participation Group, University of Tokyo, University of Utah, Vanderbilt University, University of Virginia, University of Washington, and Yale University.

\bibliography{bibliography}

\end{document}